\documentclass{ptephy_v1}

\usepackage{bm}
\usepackage{ulem}

\renewcommand\sout{\bgroup \color{red} \ULdepth=-.5ex \ULset}
\graphicspath{{./Figs/}}
\newcommand{\average}[1]{\ensuremath{\langle#1\rangle}}

\newcommand{\TCF}{C_{12}}

\newcommand{\vk}{{\bm{k}}}
\newcommand{\pik}{\pi_\vk}
\newcommand{\phik}{\phi_\vk}
\newcommand{\bOmega}{\overline{\Omega}}
\newcommand{\bomega}{\overline{\omega}}

\newcommand{\bomegak}{\bomega_\vk}

\newcommand{\omegak}{\omega_\vk}
\newcommand{\gambar}{\overline{\gamma}}
\newcommand{\gamc}{\gamma_c}

\preprintnumber{KUNS-2757/YITP-19-24} 

\begin{document}

\title{Shear viscosity of massless classical fields in scalar theory}

\author[1]{Hidefumi Matsuda}
\affil{Department of Physics, Faculty of Science, Kyoto University,
Kyoto 606-8502, Japan}
\author[2]{Teiji Kunihiro}
\affil{Yukawa Institute for Theoretical Physics, Kyoto University,
Kyoto 606-8502, Japan}
\author[2]{Akira Ohnishi}
\author[3]{Toru T. Takahashi}
\affil{National Institute of Technology, Gunma College,
Gunma 371-8530, Japan}


\begin{abstract}%
We investigate the shear viscosity of massless classical scalar fields
in the $\phi^4$ theory
on a lattice by using the Green-Kubo formula.
Based on the scaling property of the classical field,
the shear viscosity is represented using a scaling function.
Equilibrium expectation value of the time-correlation function
of the energy-momentum tensor is evaluated
as the ensemble average of the classical field configurations,
whose time evolution is obtained by solving the classical equation of motion
starting from the initial condition in thermal equilibrium.
It is found that there are two distinct damping time scales
in the time-correlation function, which
is found to show damped oscillation behavior in the early stage around
a slow monotonous decay with an exponential form,
and the slow decay part is found to dominate the shear viscosity
in the massless classical field theory.
This kind of slow decay is also known to exist in the molecular dynamics
simulation, then it may be a generic feature of dense matter.
\end{abstract}

\subjectindex{A51, D28}

\maketitle

\section{Introduction}
Classical fields have been proven useful
in describing various physical systems which have
actually quantum nature, such as
the condensate of Bose gases~\cite{GP},
inflation in the early universe~\cite{inflation},
nuclear structure~\cite{Serot:1984ey},
and gluon fields in heavy-ion collisions~\cite{McLerran:1993ni,Gelis}.
They can be used to obtain the ground states,
for example, of Bose gases by using the Gross-Pitaevskii
equation~\cite{GP}
and those of nuclei in the relativistic mean field theory~\cite{Serot:1984ey}.

Classical field theory can be also utilized
to understand the dynamical evolution
of the system~\cite{inflation,McLerran:1993ni,Gelis}.
In dynamical evolution of a quantum system,
particles can be created by a coupling with the classical field,
and its creation and their mutual collisions
produce entropy and lead to eventual thermalization.
Furthermore, it has been shown that the dynamics of the classical field
itself is responsible for entropy production in a semiclassical approximation
and thus play a significant role in the realization of an approximate
local thermal equilibrium so that
the hydrodynamical description of the time-evolution afterward becomes valid
in the case of Yang-Mills field~\cite{CYMtherm}.

The classical field dynamics and its possible relevance to
entropy production is considered as one of the key ingredients
for a hydrodynamical description of a quantum system, then it would be
intriguing to explore the properties of nonequilibrium processes
as well as equilibrium statistical physics of classical fields.
As far as we are aware of, it seems that
systematic investigations of nonequilibrium properties
such as the transport coefficients of classical fields
are limited~\cite{Homor,Aarts,Aarts:2001yn}.
Taking account of the possible role of classical fields in thermalization,
it is natural to expect that the classical field has
its own viscosity~\cite{Homor}.
The Green-Kubo formula, shown later in Eq.~\eqref{Eq:GK},
tells us that the shear viscosity can be obtained
from the integral of the time-correlation function
of the energy-momentum tensor, $\TCF(t)$ to $t=\infty$.
Since the Green-Kubo formula is obtained from the linear response theory
and should be valid independent of the theoretical framework,
it is possible to obtain the shear viscosity provided by classical fields
from the thermal average of $\TCF(t)$.

In this work, we discuss the shear viscosity of the classical field
in the massless $\phi^4$ theory using the Green-Kubo formula.
In order to make thermalized initial states,
we utilize the Langevin equation composed of streaming, diffusion
and stochastic force (latter two are collectively called the Langevin terms)
to generate configurations
in thermal equilibrium;
the achievement of thermal equilibrium is checked by examining
the equipartition of energy.
Once the thermal equilibrium is realized,
the Langevin terms are switched off
and the time evolution of the system is solely governed
by the intrinsic dynamics of the $\phi^4$ theory.
Then we calculate $\TCF(t)$
and obtain the shear viscosity of the $\phi^4$ field.
It should be noted that the $\phi^4$ theory has the scaling property,
which leads to several interesting consequences.
For example, $\TCF(t)$ and the shear viscosity are represented
by some scaling functions as shown later.

It is found that $\TCF(t)$ in massless $\phi^4$ theory has
a slow exponential decay part,
which is not found in the massive case~\cite{Homor}. 
Due to the existence of such a slow exponential decay, the shear viscosity of the massless classical scalar field is much larger than that in the massive case~\cite{Homor}.
In reality thermal fluctuations induce a thermal mass
also in the \textit{massless} $\phi^4$ theory with vanishing bare mass,
while the thermal mass squared is proportional to the coupling
and small in the weak coupling region,
where the exponential tail of $\TCF(t)$ is found to grow.
Furthermore, the massless $\phi^4$ theory would be more relevant
to the classical Yang-Mills theory, which also has vanishing bare mass and the scaling property, and has been utilized to describe the initial stage of high-energy heavy-ion collisions.
The study of the classical massless $\phi^4$ theory actually provides
insight into the classical Yang-Mills field dynamics~\cite{fluctuation}.

One of the purposes of the present work is
a detailed analysis of the time correlation of the classical
energy-momentum tensor $\TCF(t)$,
which turns out to exhibit rich phenomena beyond na\"ive expectation
such as a simple damped oscillation, especially
in the case of the massless $\phi^4$ theory.
In fact, such an analysis is necessary to extract the viscosity in
a physically meaningful way. It contains two main components;
a fast damped oscillation seen in the early stage
and a slow monotonous decay with an exponential form.
We also make a detailed analysis of the spectral function which contain
interesting physics of the excitation modes of the system.
The decay rates of these two components are
different by two or three orders in magnitude,
and the slow decay part dominates the integral of $\TCF(t)$.
We show that subtraction of the exponential term
is necessary to make the integration to $t=\infty$
in the Green-Kubo formula well-defined.
The shear viscosity is then obtained from the integral of $\TCF(t)$,
and the scaling function for the shear viscosity is determined.

One may recall that there is a subtlety
in discussing thermalization and equilibrium using classical fields.
Classical field theory can describe the low momentum modes
of quantum systems.
In the equilibrium of classical fields,
an approximate equipartition of energy holds
and each mode has the energy comparable to the temperature.
Thus the distribution of massless modes obeys the the Rayleigh-Jeans law,
which agrees with the Bose-Einstein distribution at low momenta,
and the energy density of the classical field is divergent
in the continuum limit.
There are several ways to avoid the Rayleigh-Jeans divergence.
One of the ways is to introduce the counter terms in the action.
The renormalizability of classical scalar fields was studied~\cite{Aarts},
where it is found that the static thermal expectation values
in the classical scalar theory are renormalizable
and the classical fields show surprisingly slow thermalization.
However, the behavior of 
higher-dimensional observables such as $C_{12}(t)$ is
dominated by hard particles of order temperature
and is sensitive to the ultraviolet cutoff in the classical field theory.
A sophisticated way to circumvent this problem is
to take account of the coupling with particles~\cite{Aarts:2001yn,Hatta:2011ky},
to include quantum statistical modification
of the classical variables~\cite{Ohnishi:1994wn},
or to introduce stochastic contribution
from the hard modes~\cite{Greiner:1996dx}
as expected from the Mori-Zwanzig projection operator
formalism~\cite{MoriZwanzig}.
It is, however, beyond the scope of the present work to treat
the classical field dynamics as an infrared effective dynamics
of a quantum field theory.
Instead we shall confine ourselves to investigating the transport properties
within the classical field theory by just introducing a finite lattice spacing.
Nevertheless we expect that our study will constitute a basic ingredient for an understanding
of the results given in full quantum simulations in future.

This paper is organized as follows.
In Sec. \ref{Sec:Theory},
we first explain
the lattice formulation of the classical field in the $\phi^4$ theory,
the classical field ensemble,
and the Green-Kubo formula for the shear viscosity.
We also discuss the Langevin terms, which promote thermalization,
and the equipartition nature in the equilibrium of the classical fields.
In Sec. \ref{Sec:Results},
we show the numerical results of the time-correlation function
of the energy-momentum tensor and its Fourier spectrum.
The shear viscosity is obtained from the $\omega \to 0$ limit
of the Fourier spectrum.
We also discuss the coupling dependence of the shear viscosity.
In Sec. \ref{Sec:Summary},
we summarize our work.

\section{Classical Scalar Field and Its Shear Viscosity}
\label{Sec:Theory}

\subsection{Classical Scalar Field Theory on Lattice}

We consider the $\phi^4$ theory, where the Lagrangian is given as
\begin{align}
\mathcal{L}=\frac12 \partial_\mu\phi\partial^\mu\phi
- \frac12 m^2 \phi^2 - \frac{\lambda}{24} \phi^4
\ .
\end{align}
On a $L^3$ lattice, the Hamiltonian is given by
\begin{align}
H=\frac{1}{2}
\sum_{\bm{x}}\left[\pi^2(x)
+ \left(\partial_i\phi(x)\right)^2+m^2 \phi(x)^2
+ \frac{\lambda}{12}\phi^4(x) \right],\nonumber\\
\end{align}
where ($\phi(x), \pi(x)=\dot{\phi}(x)$) are the canonical variables
and $\partial_i$ denotes the forward difference operator
in the $i$-th direction,
i.e. $\partial_i\phi(x)=\phi(x+\hat{e}^i)-\phi(x)$.
We take all quantities normalized by the lattice spacing $a$
throughout this article.

We discuss the evolution of the scalar field in the classical field
approximation, where the field variables $(\phi(x),\pi(x))$ are regarded
as $c$-numbers.
Instead of taking account of fluctuations of the field operators
in the evolution in each classical field configuration,
statistical fluctuations are taken into account
by using the classical field ensemble as described later.
Then the above Hamiltonian is regarded as a classical Hamiltonian,
which gives
the classical equation of motion,
\begin{align}
\dot{\phi}(x) = \frac{\partial H}{\partial \pi(x)}\ ,\quad
\dot{\pi}(x) = -\frac{\partial H}{\partial \phi(x)}.
\label{Eq:EOM}
\end{align}
We define the off-diagonal matrix elements of the energy-momentum tensor
of scalar fields on the lattice as
\begin{eqnarray}
T_{ij}&\equiv&\left(\partial^{\rm c}_i\phi(x)\right)\left(\partial^{\rm c}_j\phi(x)\right)
\quad (i \not= j)\ .
\end{eqnarray}
Here we adopt the central difference,
$\partial^{\rm c}_i\phi(x)
=\left[\phi(x+\hat{e}^i)-\phi(x-\hat{e}^i)\right]/2$.
With this prescription,
$\partial^{\rm c}_i\phi(x)$ is located in the same space-time point
as $\phi(x)$ and the matrix element becomes more symmetric
in the space directions.
Then the central difference $\partial^{\rm c}_i$ is found
to give a better definition of the energy-momentum tensor
than the forward difference operator $\partial_i$.

\subsection{Green-Kubo Relation}
We evaluate the shear viscosity of the classical field
by using the Green-Kubo formula which is a powerful relation
in the linear response theory~\cite{zubarev1996statistical}.
\begin{align}
&\eta = \lim_{\omega \to 0}
\frac{1}{T}\int^\infty_0 dt\, e^{i\omega t} \TCF(t),\label{Eq:GK}\\
&\TCF(t) = V \average{\tau_{12}(t)\tau_{12}(0)}_{\rm eq},\label{Eq:C12}\\
&\tau_{12}(t) \equiv \frac{1}{V}\int d^3x \tau_{12}(\bm{x}, t),
\end{align}
where $T$ and $V$ denote temperature and volume, respectively,
$\tau_{12}(t)$ is the space-averaged off-diagonal matrix element
of the energy-momentum tensor,
and $\langle \cdots \rangle_\mathrm{eq}$ represents the expectation value
in equilibrium.

In the long time-difference limit $t\to\infty$,
the correlation disappear and $\TCF(t)$ is expected to vanish.
\begin{eqnarray*}
\TCF(t)=
V\,\average{\tau_{12}(t)\tau_{12}(0)}_{\rm eq}\xrightarrow{t\to\infty}
V\,\average{\tau_{12}(0)}^2_{\rm eq}=0.
\end{eqnarray*}
The final equality is expected from the isotropy of thermal systems.
Thus when the correlation disappears faster than $1/t$,
the integral in Eq.~\eqref{Eq:GK} converges to a finite value.
In actual calculations, the correlation has a long-time tail
and it is not easy to obtain the converged integral.
As demonstrated later,
we find that it is possible to evaluate the integral to infinity safely
by subtracting the slowly decreasing exponential function
and adding the exponential contribution separately.

\subsection{Equilibrium in Classical Field Ensemble}

In the Green-Kubo formula, we need the equilibrium expectation value of
the time-correlation function of the energy-momentum tensor.
We here obtain the equilibrium expectation value
by using configurations in the classical field ensemble.
We prepare $N_\mathrm{conf}$ initial classical field configurations
in equilibrium.
Then we regard the average of observables in this ensemble
as the expectation values in classical equilibrium,
\begin{align}
\average{ \mathcal{O} }_{\rm eq}
\simeq& \frac{1}{N_\mathrm{conf}}\sum^{N_\mathrm{conf}}_{i=1} \mathcal{O}_i,
\end{align}
where $\average{ \mathcal{O}}_{\rm eq}$ is the expectation value
of the observable $\hat{ \mathcal{O}}$,
$N_\mathrm{conf}$ is the number of classical field configurations,
and $\mathcal{O}_i$ is the observable in the $i$-th configuration.

In order to obtain many equilibrium classical field configurations efficiently,
we introduce the artificial Langevin terms in the thermalization stage.
In the $\phi^4$ theory, it takes a long time for the system
to equilibrate especially at low $T$ or $\lambda$,
then a large computational resource is required
to prepare a large number of equilibrium configurations.
By introducing the Langevin terms, thermalization is hastened.
We add the Langevin terms, combination of the diffusion and stochastic forces,
to the right hand side of the equation of motion
for $\pi(x)$,
\begin{eqnarray}
\dot{\pi}(x) &=& -\frac{\partial H}{\partial \phi(x)} - \gamma \pi(x) + R(x),
\label{Eq:Langevin}
\end{eqnarray}
where $\gamma$ is the diffusion coefficient
and $R(x)$ represents the fluctuation,
which is assumed to obey a Gaussian distribution function,
namely the white noise.
The strength of the fluctuation is determined by the Einstein relation,
\begin{align}
\average{R(x)R(x')}=&2\gamma T\delta(t-t')\delta_{\bm{x},\bm{x}'},
\end{align}
where $T$ is the temperature.

When equilibrium is reached, classical field configurations are
distributed according to the Boltzmann weight.
The partition function of the classical field on the lattice is given as
\begin{align}
\mathcal{Z}
= \int \mathcal{D}\pi \mathcal{D}\phi \exp(-H/T)
=& \int \mathcal{D}\pi \exp\left(-\frac{1}{2T} \sum_x \pi^2(\bm{x})\right)
\int \mathcal{D}\phi\,e^{-H_\phi/T}
\nonumber\\
=&\prod_{\bm{k}}\left[
\int d\pi(\bm{k}) e^{-|\pi(\bm{k})|^2/2T}
\right]
\int \mathcal{D}\phi\,e^{-H_\phi/T}
\ ,
\end{align}
where $\bm{k}$ denotes the lattice momentum $\bm{k}=2\pi(n_x,n_y,n_z)/L$
($n_i=0, 1, \cdots L-1$),
$\pi(\bm{k})=\sum_{\bm{x}}\pi(\bm{x})e^{-i\bm{k}\cdot\bm{x}}/\sqrt{L^3}$
is the Fourier transform of $\pi(\bm{x})$,
and $H_\phi$ is a part of the Hamiltonian containing only $\phi$.

Note that $\pi(\bm{k})$ is complex in general and
$\pi(\bm{k})$ and $\pi(-\bm{k})$
satisfies $\pi(-\bm{k})=\pi(\bm{k})^*$, then
the integral should be regarded to be
$d\pi(\bm{k})d\pi(-\bm{k})=
2 d\,\mathrm{Re}\pi(\bm{k})\,d\,\mathrm{Im}\pi(\bm{k})$.
Only in those cases with all the momentum components being $0$ or $\pi$,
$\pi(\bm{k})$ is real.

Since the Hamiltonian is quadratic in $\pi(\bm{x})$ as well as in $\pi(\bm{k})$,
the expectation values in equilibrium should be given as
\begin{align}
\average{\pi(\bm{x})^2}_\mathrm{eq}=T
,\
\average{|\pi(\bm{k})|^2}_\mathrm{eq}=T
.
\label{Eq:pisqav}
\end{align}
Thus we can measure the temperature
of classical equilibrium system from the expectation values
of $\pi(\bm{x})^2$ or $|\pi(\bm{k})|^2$,
where the expectation value does not depend on the position and the momentum.
In section \ref{sbEP}, we check this relation.

\subsection{Scaling property}

In actual calculations,
we concentrate on the time evolution of classical fields,
where equation of motion has the scaling property:
The equation of motion, Eqs.~\eqref{Eq:EOM},
is invariant under the transformation,
\begin{align}
\phi \to \phi/\sqrt{\alpha} ,\
\pi \to \pi/\sqrt{\alpha} ,\
\lambda \to \alpha\lambda .
\end{align}
Then the phase space trajectories are the same with this transformation,
which is referred to as the $\lambda$ deformation hereafter.

The $\lambda$ deformation results in
$H_{\lambda} \to H_{\alpha\lambda}/\alpha$,
with $H_\lambda$ being the Hamiltonian with the coupling $\lambda$,
and the temperature defined above can be regarded as
\begin{align}
T_{\lambda}
=\langle \pi_x^2 \rangle_{\lambda}
=\alpha \langle \pi_x^2 \rangle_{\alpha\lambda}
=\alpha T_{\alpha\lambda} .
\end{align}

By using the $\lambda$ deformation,
the time-correlation function and the shear viscosity
at a given $(\lambda, T)$ are found to be related
with those having the same $\lambda T$ value as,
\begin{align}
\TCF(t;\lambda,T)=&\alpha^2\TCF(t; \alpha\lambda,T/\alpha)
\ ,\\
\eta(\lambda,T)=&\alpha\eta(\alpha\lambda,T/\alpha)
\ .
\end{align}
Then it is enough to vary $\lambda$ and to keep $T$ fixed in order to
evaluate $\TCF(t)$ and $\eta$ at various combinations of $\lambda$ and $T$.
These relations tantamount to the existence of the scaling functions
which determine $\TCF(t;\lambda,T)$ and $\eta(\lambda,T)$,
\begin{align}
\lambda^2\TCF(t;\lambda,T)=&f_C(t;\lambda T)\ ,\\
\lambda^2T\eta(\lambda,T)=&f_\eta(\lambda T)\ .
\label{Eq:feta}
\end{align}
It should be remembered that these scaling functions depend also on $m^2$.

\section{Numerical results}
\label{Sec:Results}

\subsection{Calculation Setup}

In actual calculations, we concentrate on the case with $m=0$.
The coupling is chosen to be $\lambda=0.5, 1, 3, 5, 10$ and $30$.
The calculations are performed with the three different lattice sizes
$L=16, 32$ and $64$,
and we shall mainly show the results with $L=32$,
although significant/insignificant lattice-size dependence will be discussed in some important cases.
Equilibrium classical field configurations are generated
by introducing the Langevin terms with $\gamma=0.1$.
Time evolution with the Langevin terms is calculated
with a fixed lattice temperature $T=1$
for a duration of $t_\mathrm{eq} = 1/\gamma$,
where we have confirmed that classical equilibrium is reached.
After preparing $N_\mathrm{conf}=1000$ classical field
equilibrium configurations for each value of $\lambda$,
we perform the time evolution calculation without the Langevin terms
until $t=10^4/\lambda^2$.
The time evolution is calculated in the leap-flog scheme
with the time step of $\Delta t=0.01$.

\subsection{Examination of Equilibrium}\label{sbEP}
We first check that classical equilibrium is reached correctly
by using the classical equilibrium relation in Eq.~\eqref{Eq:pisqav}
at various frequencies.
Thermalization process with the Langevin terms is discussed
in Appendices \ref{App:TCFLang} and \ref{App:TCFLangED}.

\begin{figure}[tbhp]
\begin{center}
\includegraphics[width=80mm,bb=0 0 360 252]{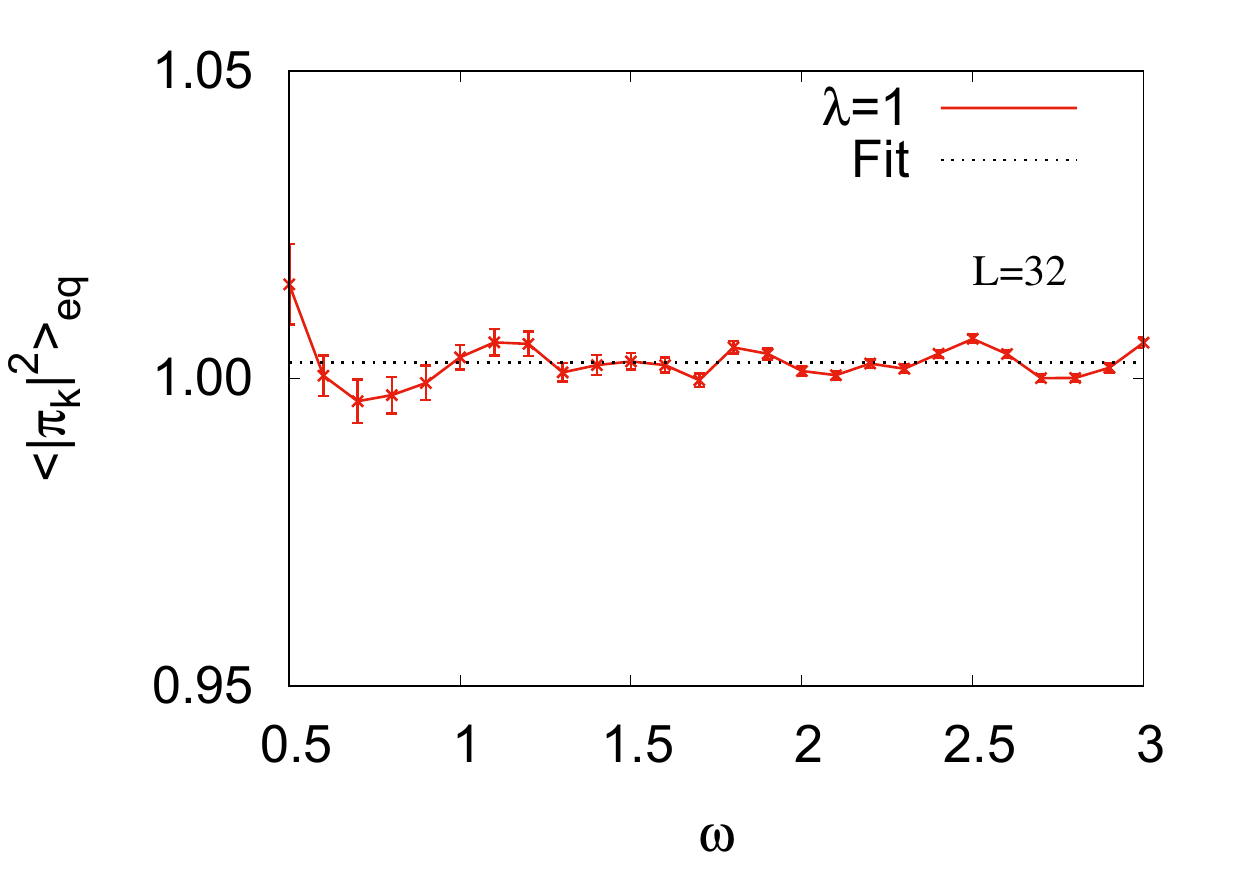}
\end{center}
\caption{Thermal expectation value of
$|\pi(t, \bm{k})|^2$, $\average{|\pi(t, \bm{k})|^2}_{\rm eq}$,
as a function of the lattice frequency $\omegak$ on the $32^3$ lattice
after the thermalization process with the Langevin terms.
}
\label{figEP}
\end{figure}

In Fig.~\ref{figEP}, we show
$\average{|\pi(t, \bm{k})|^2}_{\rm eq}$ as a function of the frequency
\begin{align}
\omegak = 2 \sqrt{\sin^2(k_1/2)+\sin^2(k_2/2)+\sin^2(k_3/2)}
\label{Eq:omegak}
\end{align}
on a $32^3$ lattice after the equilibration process with the Langevin terms.
We find that $\average{|\pi(t, \bm{k})|^2}_{\rm eq}$ is almost constant
and is close to the given temperature, $T=1$.
By fitting a constant $T_\mathrm{fit}$ to $\average{|\pi(t, \bm{k})|^2}_{\rm eq}$, we obtain,
\begin{eqnarray}
T_{\rm fit} = 1.002 \pm 0.007 .
\end{eqnarray}
Thus the classical equilibrium is confirmed to be reached,
where the measured temperature is consistent
with the target one within the relative error of 0.2 \%.

\subsection{Time-correlation function of energy-momentum tensor}

\begin{figure}[tbhp]
\begin{center}
\includegraphics[width=80mm,bb=0 0 360 252]{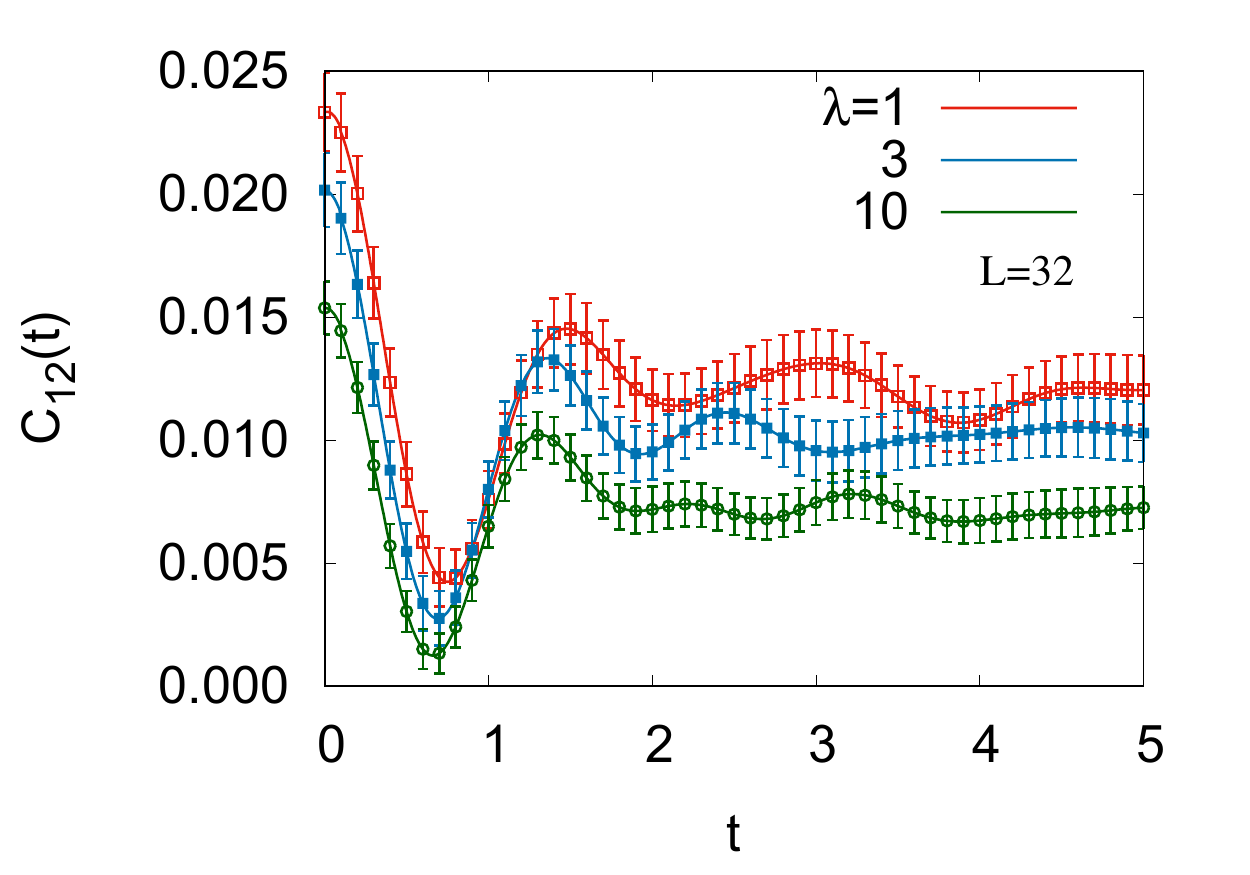}
\end{center}
\caption{
The short time behavior ($t<5$) of
the time-correlation function of the energy momentum tensor,
$\TCF(t)=V\average{\tau_{12}(t)\tau_{12}(0)}$,
on the $32^3$ lattice with the coupling $\lambda=1, 3$ and $10$.}
\label{Fig:TCFshort}
\end{figure}

We shall now discuss the time-correlation function $\TCF(t)$
of the energy-momentum tensor in the $\phi^4$ theory.
In Fig.~\ref{Fig:TCFshort},
we show the time-correlation function of the energy-momentum tensor
for several values of $\lambda$ in the short time period, $t < 5$.
In the time region of $t < 5$,
the correlation function seems to show damped oscillation
around a constant value.
The oscillation period is around $\tau_\mathrm{short}=1.2-1.4$.
Then the frequency is evaluated as
$\omega_\mathrm{short}=2\pi/\tau_\mathrm{short} \sim (4.5-5.2)$.

\begin{figure}[tbhp]
\begin{center}
\includegraphics[width=80mm,bb=0 0 360 252]{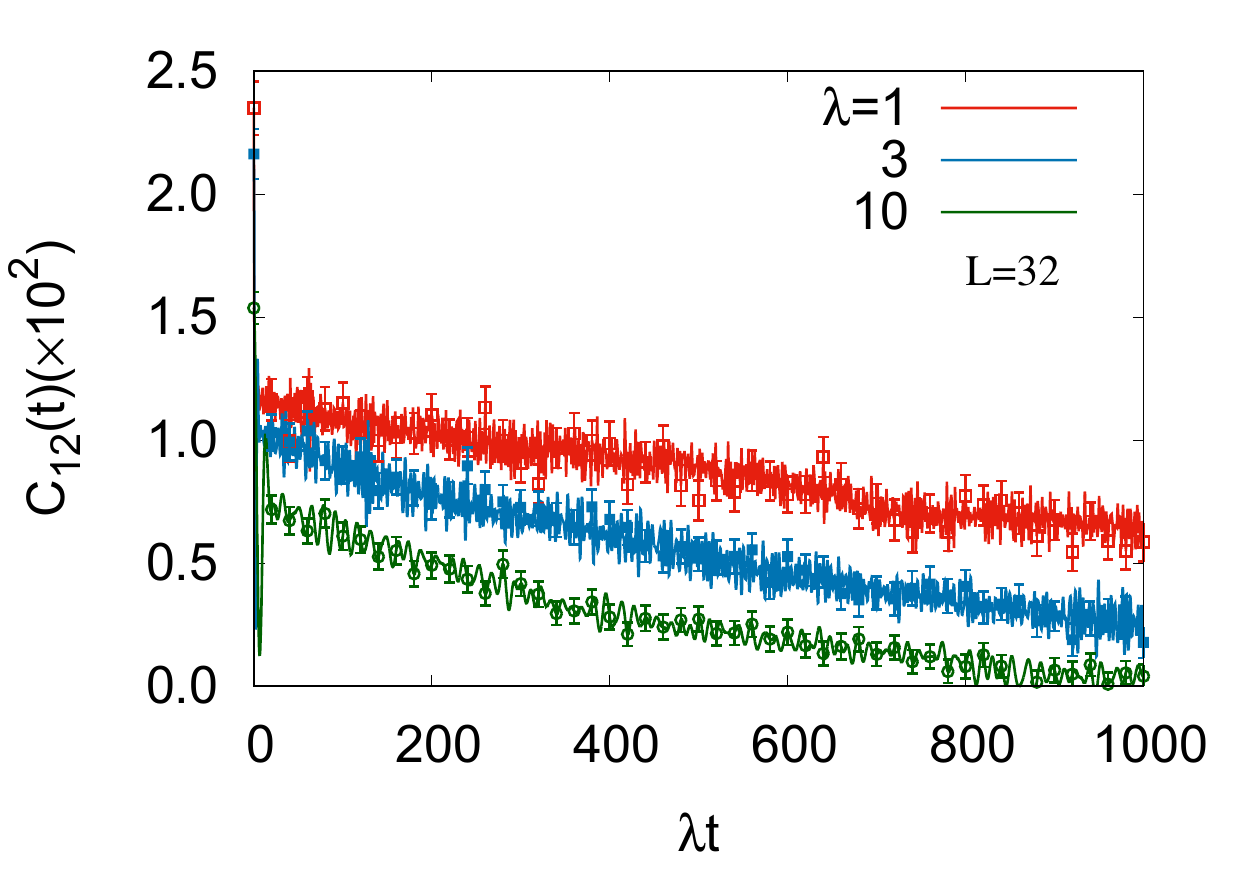}
\\
\includegraphics[width=80mm,bb=0 0 360 252]{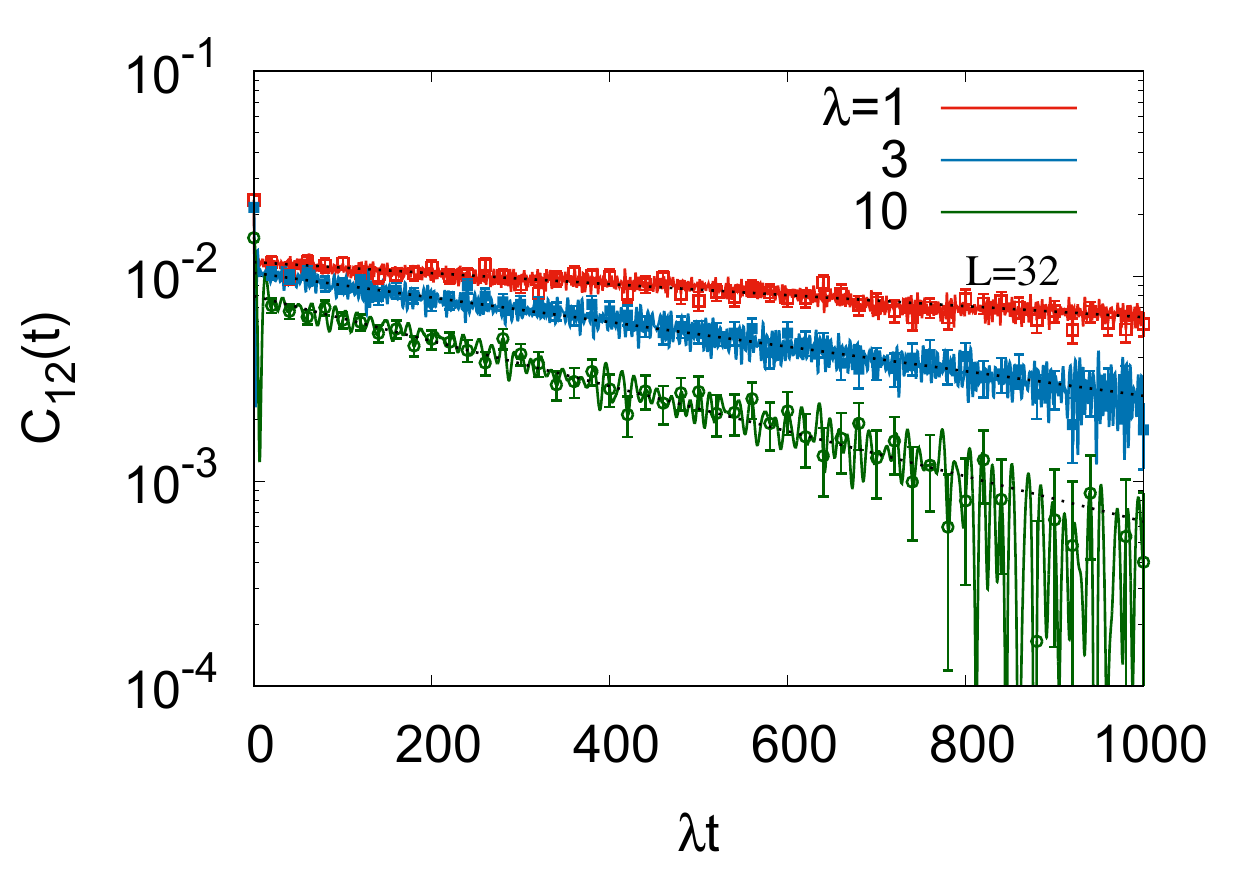}
\end{center}
\caption{The same as Fig.~\ref{Fig:TCFshort} but results in the long time range,
$\lambda t \leq 1000$, are shown.
In the upper and lower panels, we show the results
in the linear and log scales, respectively.
Black dotted lines show the single-exponential functions,
which fit the long time behavior.}
\label{Fig:TCFlong}
\end{figure}

\begin{figure}[tbhp]
\begin{center}
\includegraphics[width=80mm,bb=0 0 360 252]{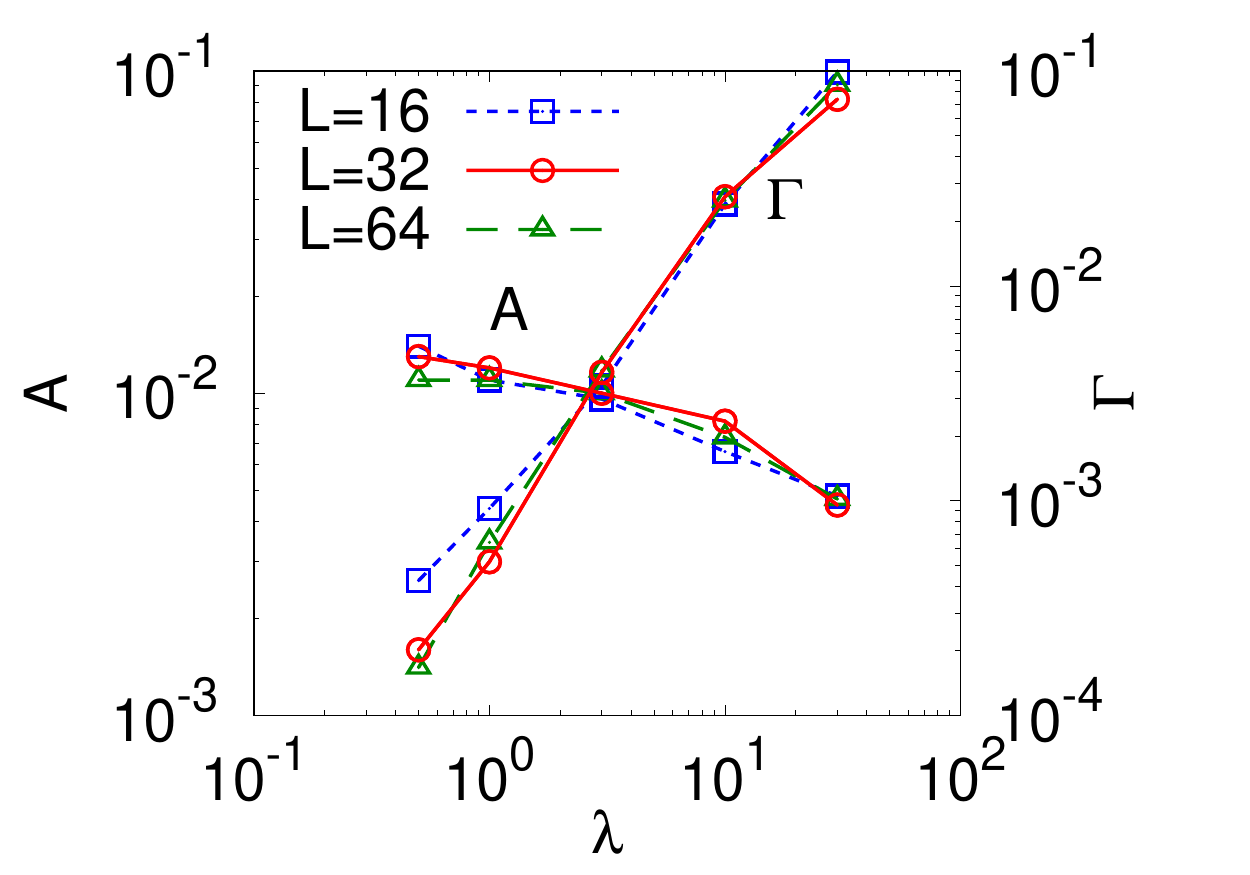}
\end{center}
\caption{The coupling dependence of $(A, \Gamma)$,
where the time-correlation function is fitted
by the exponential function, $\TCF(t)\simeq A\exp(-\Gamma t)$.}
\label{Fig:AGam}
\end{figure}

In contrast,
longer time results show that the background decreases with a larger rate for larger $\lambda$, as shown in Fig.~\ref{Fig:TCFlong}.
We find that
the correlation is found to oscillate around a slow monotonous decay.
This slow decay may be approximated by an exponential function,
\begin{align}
\TCF^\mathrm{(exp)}(t)=A \exp(-\Gamma t)
\ .\label{Eq:decay}
\end{align}
We show the fitted results using one exponential function
by black dotted lines in Fig.~\ref{Fig:TCFlong},
whose parameters are summarized in Table~\ref{Tab:eta},
and their $\lambda$ dependence is shown in Fig.~\ref{Fig:AGam}.
The fitting parameter $A$ is not very sensitive to $\lambda$,
while the $\lambda$ dependence of $\Gamma$ is strong.
When we add a constant or another exponential term,
$\chi^2$ is slightly improved
but the correlation among parameters
makes parameter uncertainties much larger.
Since $\TCF(t)$ should approach zero in the long-time limit,
$t \to \infty$, we assume that the decay is assumed to be given
by the single-exponential function in the later discussions.

\begin{table}[htbp]
\caption{Parameters in the exponential decay part of $\TCF(t)$
($A$ and $\Gamma$) and the scaling function($f_\eta(\lambda T)=\lambda^2T\eta$)
at $\lambda=0.5, 1, 3, 10$ and $30$
on the $16^3, 32^3$ and $64^3$ lattices.
We show the results at $T=1$.}
\label{Tab:eta}
\begin{center}
\begin{tabular}{rr|ccc}
\hline\hline
$\lambda$ & $L$ & $A$ & $\Gamma/\lambda^2$ &$f_\eta(\lambda T)$\\
\hline
0.5& 16 & 14$\times10^{-3}$ & 17$\times10^{-4}$ &8.4\\
& 32 & 13$\times10^{-3}$ & 8.1$\times10^{-4}$ & 16 \\
& 64 & 11$\times10^{-3}$ & 6.7$\times10^{-4}$ & 17 \\
\hline
1& 16 & 11$\times10^{-3}$ & 9.2$\times10^{-4}$ & 12 \\
& 32 & 12$\times10^{-3}$ & 5.2$\times10^{-4}$ & 22 \\
& 64 & 11$\times10^{-3}$ & 6.4$\times10^{-4}$ & 17 \\
\hline
3& 16 & 9.6$\times10^{-3}$ & 3.8$\times10^{-4}$ & 25\\
& 32 & 10$\times10^{-3}$ & 4.4$\times10^{-4} $ & 23 \\
& 64 & 10$\times10^{-3}$ & 4.5$\times10^{-4}$ & 23 \\
\hline
10& 16 & 6.6$\times10^{-3}$ & 2.4$\times10^{-4}$& 27 \\
& 32 & 8.2$\times10^{-3}$ & 2.6$\times10^{-4}$ & 31 \\
& 64 & 7.3$\times10^{-3}$ & 2.5$\times10^{-4}$ & 29 \\
\hline
30& 16 & 4.8$\times10^{-3}$ & 1.1$\times10^{-4}$& 43 \\
& 32 & 4.5$\times10^{-3}$ & 0.82$\times10^{-4}$ & 54 \\
& 64 & 4.7$\times10^{-3}$ & 0.96$\times10^{-4}$ & 49 \\
\hline\hline
\end{tabular}
\end{center}
\end{table}

\subsection{Fourier spectrum}

Next, let us discuss the Fourier spectrum $\rho(\omega)$
of the time-correlation function of the energy-momentum tensor, $\TCF(t)$,
\begin{align}
\rho(\omega)
=&\mathrm{Re}\int_0^\infty dt e^{i\omega t} \TCF(t)
\nonumber\\
=&\frac12\left[
\int_0^\infty dt e^{i\omega t} \TCF(t)
+
\int_{-\infty}^0 dt e^{i\omega t} \TCF(-t)
\right]
\nonumber\\
\simeq & \lim_{N\to \infty, \Delta t\to0}\frac{\Delta t}{2} \sum_{n=-N/2+1}^{N/2} e^{i\omega n \Delta t} \TCF(|n \Delta t|)
,\label{Eq:Fourier}
\end{align}
where $\Delta t=2t_\mathrm{max}/N$.
The last line in Eq.~\eqref{Eq:Fourier} is nothing but the discrete
Fourier transformation, and we obtain the value of $\rho(\omega)$
at $\omega=n\pi/t_\mathrm{max}$.
The zero frequency limit $\lim_{\omega \to 0} \rho(\omega)$
is equal to the integral of $\TCF(t)$ upto $t=\infty$,
and is related with the shear viscosity as given in Eq.~\eqref{Eq:etarho}.
The relation between the Fourier spectrum and the spectral density
is described in Appendix~\ref{App:SD}.

\begin{figure}[htbp]
\begin{center}
\includegraphics[width=80mm,bb=0 0 360 252]{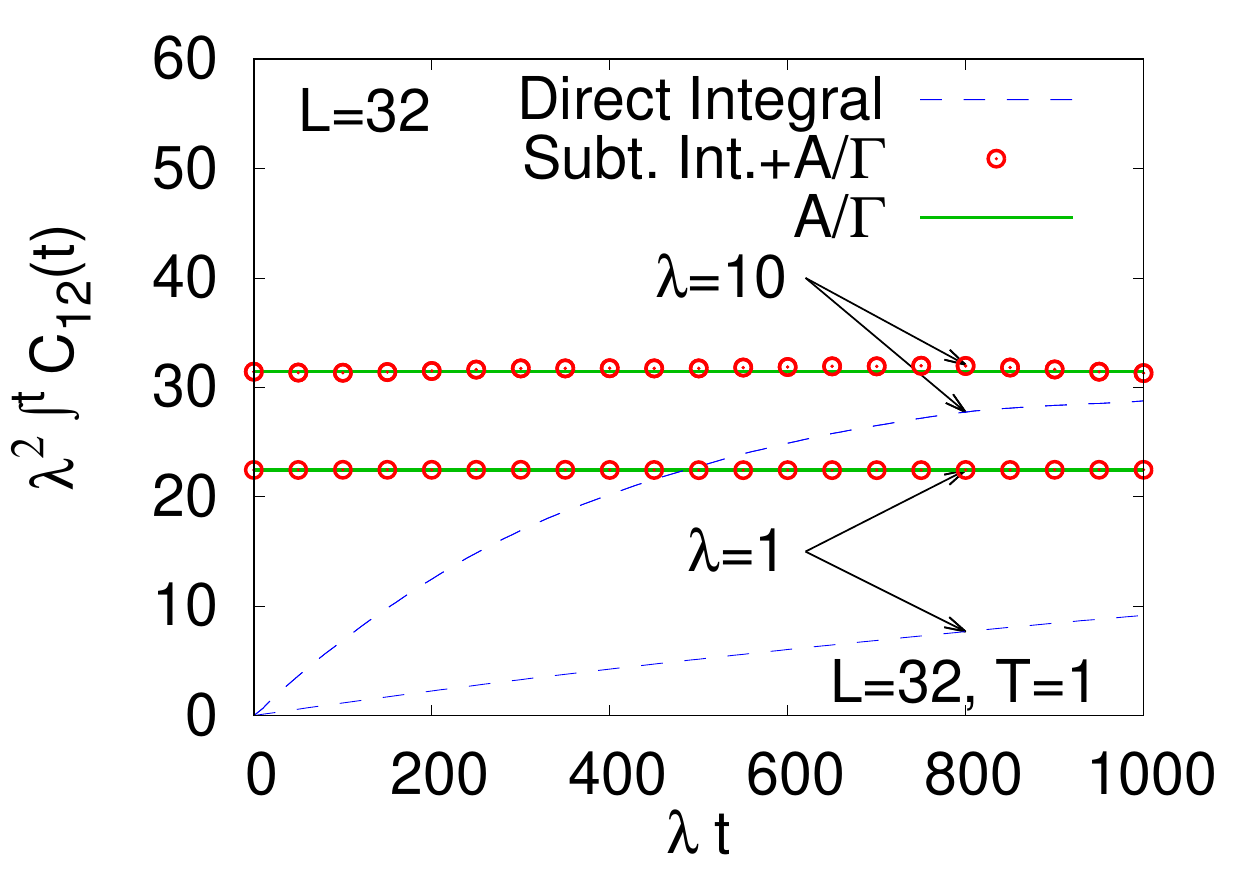}
\end{center}
\caption{Convergence of the integral of the time-correlation function
at $\lambda=1$ and $10$ on the $32^3$ lattice.
The blue dashed lines show the results of a naive integration up to $t$,
while the open circles show the sum of the integrals up to $t$ of the subtracted correlation function,
$\TCF^\mathrm{(sub)}(t)=\TCF(t)-\TCF^{\mathrm{(exp)}}(t)$
and the integral upto $t\to\infty$ of the
exponential-decay function $A\exp(-\Gamma t)$, which is equal to $A/\Gamma$.
The solid lines show $A/\Gamma$.}
\label{Fig:conv}
\end{figure}

Integration up to $t=\infty$ turns out to be rather involved because of the exponential but slow decay
of $\TCF(t)$ at large $t$.
As an example, we show the integral of $\TCF(t)$ to $t$
in Fig.~\ref{Fig:conv}.
At $\lambda=1$, the decay rate $\Gamma\simeq 5\times 10^{-4}$ is small,
and the integral is not yet saturated at $\lambda t=1000$
as shown by the lower dashed curve.
At $\lambda=10$, the integral is convergent but is still increasing
at $\lambda t=1000$.
In order to avoid these problems,
we first subtract the exponential decay part from $\TCF(t)$,
perform the integration,
and add the Fourier transformation of the exponential decay part, which is identically equal to $A/\Gamma$.
With this separation, the integral to $t\to\infty$ is obtained
efficiently for not only large but also small $t$
as shown by the open circles in Fig.~\ref{Fig:conv}.

The Fourier transformation is performed in a similar manner.
We make the Fourier transformation of
the subtracted time-correlation function $\TCF^\mathrm{(sub)}(t)$,
and the Fourier transformation of the exponential decay part
is added after the discrete Fourier transformation
in the third line of Eq.~\eqref{Eq:Fourier},
\begin{align}
\rho(\omega)
=&\frac{1}{2}\mathrm{F.T.}[\TCF(t)]
=\frac{1}{2}\left\{
\mathrm{F.T.}[\TCF^\mathrm{(sub)}(t)]
+\mathrm{F.T.}[\TCF^\mathrm{(exp)}(|t|)]
\right\} ,\\
\TCF^\mathrm{(sub)}(t)=&\TCF(t)-\TCF^\mathrm{(exp)}(|t|) .
\end{align}
The contribution of the exponential function Eq.~\eqref{Eq:decay}
to the Fourier spectrum takes a form of a Lorentzian,
\begin{align}
\rho^\mathrm{(exp)}(\omega)
=& \frac12 \mathrm{F.T.}[\TCF^\mathrm{(exp)}(|t|)]
=\frac{A\Gamma}{\Gamma^2+\omega^2}
\ .
\end{align}

\begin{figure}[tbhp]
\begin{center}
\includegraphics[width=80mm,bb=0 0 360 252]{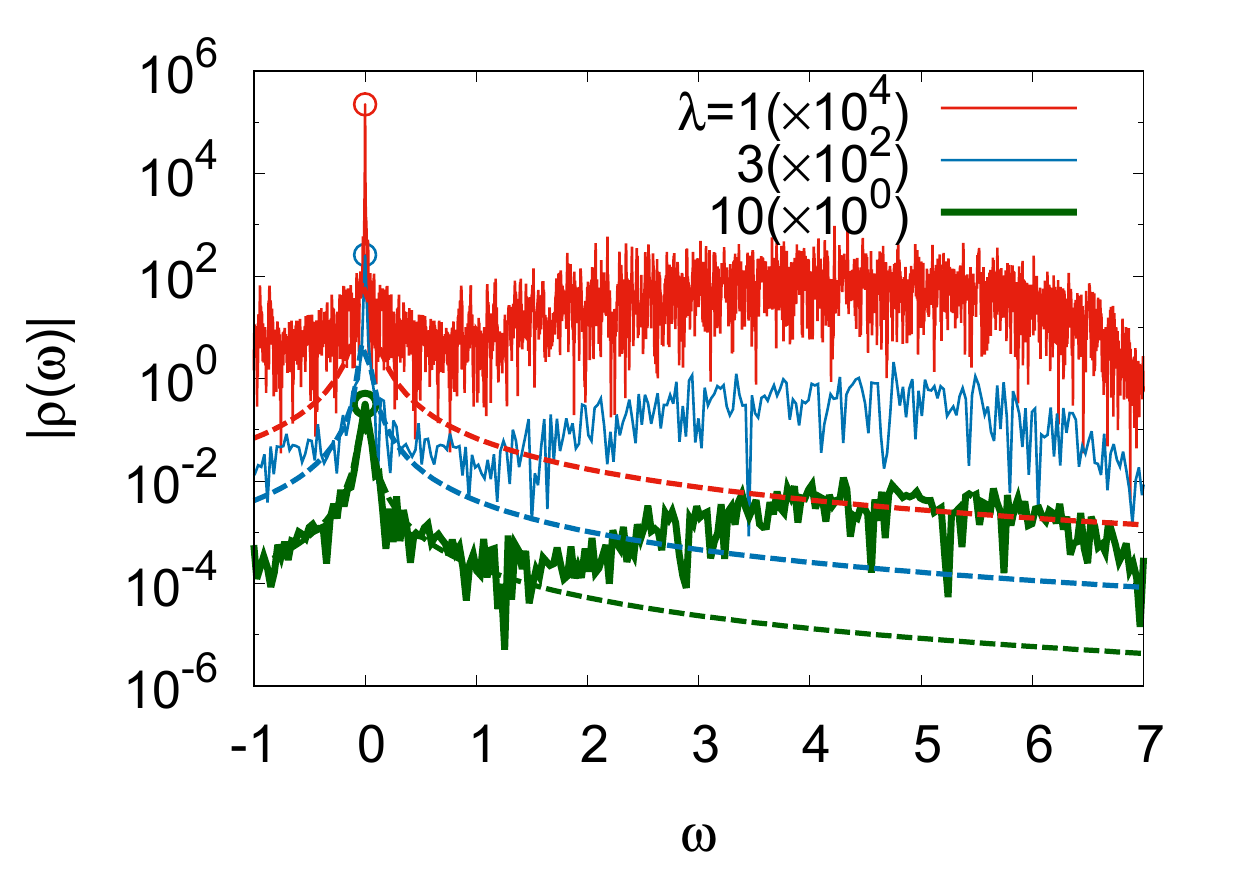}
\\
\includegraphics[width=80mm,bb=0 0 360 252]{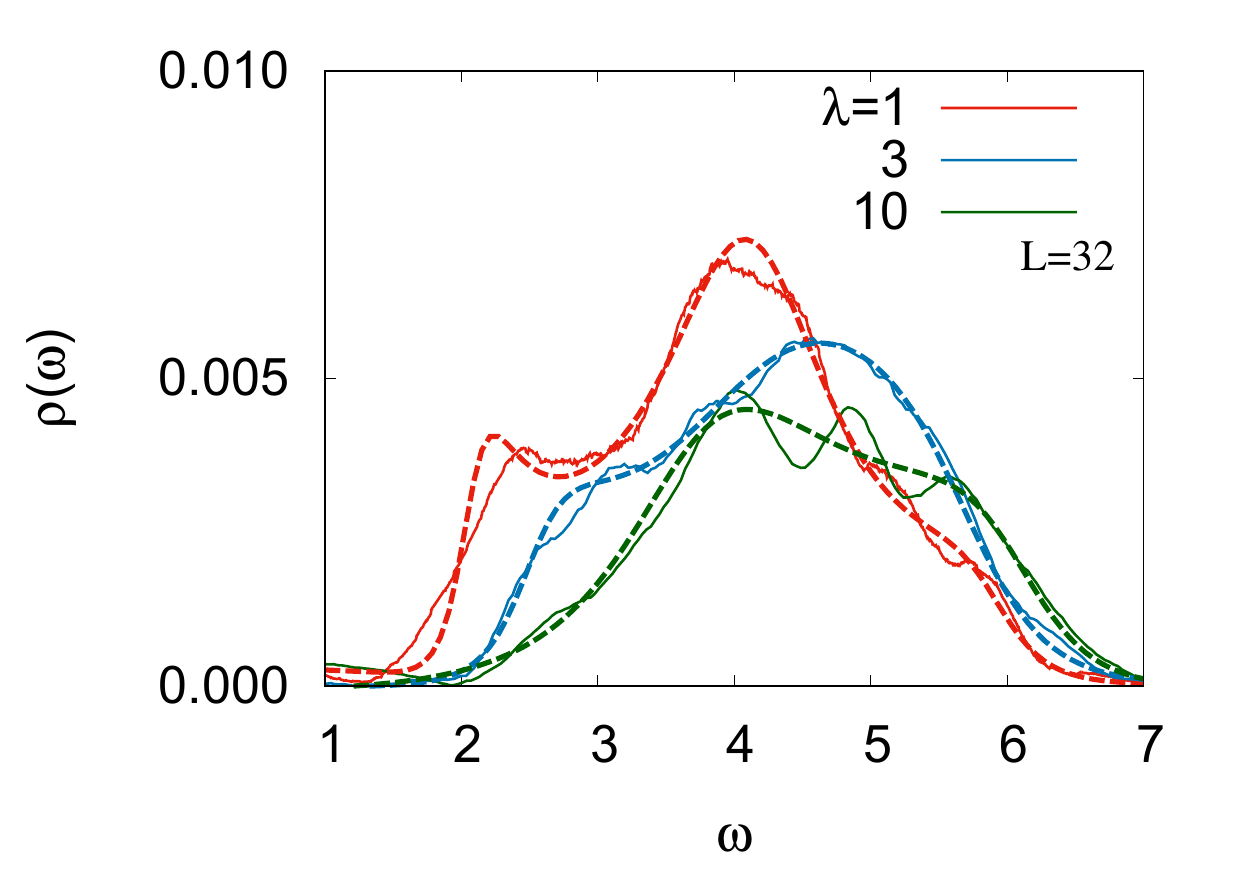}
\end{center}
\caption{Fourier spectra of the time-correlation function, $\rho(\omega)$,
on the $32^3$ lattice with the coupling
$\lambda=1$ (red), $3$ (green) and $10$ (blue).
In the upper panel, we show $\rho(\omega)$ in the log scale.
Thin dashed curves show the Lorentzian functions,
which are the Fourier transformations of the exponential decay parts.
Thick dashed lines show the Fourier spectra,
smeared by a Gaussian function
with the standard deviation of $\Delta\omega=0.2$.
In the lower panel, we show $\rho(\omega)$ in the high frequency region,
$\omega \geq 1$.
The smeared Fourier spectra of the original $\TCF(t)$ are shown by solid curves.
Dotted lines show the Fourier transform of the time-correlation function
in the early stage,
which are fitted and represented by three damped oscillators.
}\label{Fig:Spec}
\end{figure}

In Fig.~\ref{Fig:Spec}, we show the Fourier spectra
at $\lambda=1, 3$ and $10$ on a $32^3$ lattice.
We find two structures, a peak at $\omega=0$
and a bump at $\omega\simeq 4$.
At small $\omega$, the Fourier spectrum is dominated by the Lorentzian function
from the slow decay part,
shown by the dotted curves in the upper panel of Fig.~\ref{Fig:Spec}.
With decreasing $\lambda$, the decay rate $\Gamma$ becomes smaller
and hence the peak becomes sharper.
The peak height $A/\Gamma$ at the zero frequency
has milder $\lambda$ dependence.

At high frequencies,
a broad bump appears around $\omega=(2-6)$.
As already mentioned, the bump around $\omega=4$
is attributed to the oscillatory behavior in the early stage.
Since this oscillatory behavior exists irrespective of the magnitude of
$\lambda$, a remnant of the oscillation may partly constitute
the high frequency bumps.
We fit the oscillation curves observed in $\TCF(t)$ in the early stage
by three damped oscillators, and the Fourier transform of them
are shown by dotted curves in the lower panel of Fig.~\ref{Fig:Spec}.
As a reference,
the spectra of the original $\TCF(t)$ are shown by solid curves,
which coincide well with the dotted curves,
the spectra of three damped oscillators.

This broad bump also appears in the free case, $\lambda=0$.
Then the frequency of $\TCF(t)$ should be given as the sum of frequencies
of two momentum modes, $\omega=\omega_{\bm{k}_1}+\omega_{\bm{k}_2}$,
in the free case.
The single mode frequency is in the range of
$0 \leq \omega_{\bm{k}}\leq 2\sqrt{3}$,
then the sum of two frequencies would be in the range
$0 \leq \omega \leq 4\sqrt{3}\simeq 7$ without the interaction,
which agrees with the spread of the bump.
The frequency difference among various two-momentum modes
causes the decoherence
and damping of the time-correlation function in the early stage.
In Appendices \ref{App:TCFLang} and \ref{App:TCFLangED},
we demonstrate that the two-momentum modes explain the damped oscillatory
behavior in the early stage in free scalar theory with Langevin terms.

\subsection{Shear viscosity}
\label{Subsec:Shear}

By using the Fourier spectrum $\rho(\omega)$,
the shear viscosity is obtained
from the low-frequency limit,
\begin{align}
\eta = \lim_{\omega \to 0} \frac{\rho(\omega)}{T}\ .
\label{Eq:etarho}
\end{align}
As already discussed, the Fourier transform of the exponential decay part
of $\TCF(t)$ appears as a Lorentzian
and dominates the Fourier spectrum at low frequencies.
The contribution of the subtracted part is less than 1 \%
of the exponential decay part on a $32^3$ lattice.
Therefore, it is sufficient to consider the Lorentzian part
in calculating the shear viscosity
with a few percent accuracy.
In this approximation, the shear viscosity is found to be
\begin{align}
&\eta = \lim_{\omega \to 0} \frac{\rho(\omega)}{T}
\simeq \lim_{\omega \to 0} \frac{\rho^\mathrm{(exp)}(\omega)}{T}
= \frac{A}{T\,\Gamma}
\ .
\end{align}
where $A$ and $\Gamma$ are the fitting parameters
introduced in Eq.~\eqref{Eq:decay}.

We now discuss the scaling function $f_\eta(\lambda T)$
given in Eq.~\eqref{Eq:feta}.
While the actual numerical calculations are performed at $T=1$,
we can evaluate the shear viscosity at various $(\lambda,T)$
by using $f_\eta$ as,
\begin{align}
\eta(\lambda,T)=\frac{f_\eta(\lambda T)}{\lambda^2T}
\ .
\end{align}
\begin{figure}[htbp]
\begin{center}
\includegraphics[width=80mm,bb=0 0 360 252]{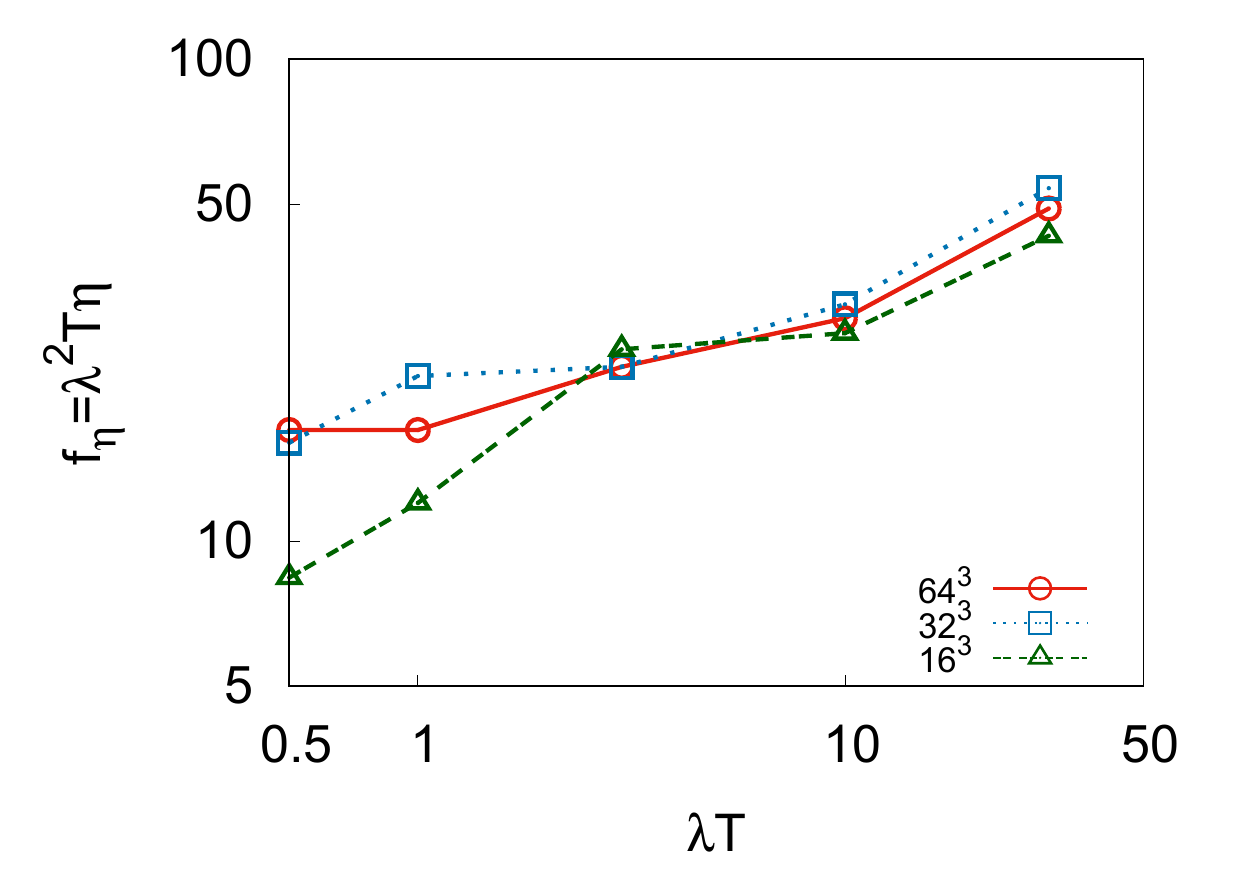}
\end{center}
\caption{The scaling function for the shear viscosity,
$f_\eta(\lambda T)=\lambda^2T\eta(\lambda,T)$
as a function of $\lambda T$.
The red, blue, green lines show the numerical results
on the $16^3, 32^3$ and $64^3$ lattices, respectively.}
\label{Fig:shear2}
\end{figure}

In Fig.~\ref{Fig:shear2}, we show the scaling function $f_\eta(\lambda T)$
as a function of $\lambda T$ on the $16^3, 32^3$ and $64^3$ lattices.
While $f_\eta$ measured on a $16^3$ lattice show decreasing behavior
at small $\lambda T$,
results on $32^3$ and $64^3$ lattices
seem to converge to a constant at small $\lambda T$,
$f_\eta(\lambda T) \simeq 17$, which implies that $\eta \propto 1/\lambda^2$ for small $\lambda T$.
Incidentally, it may be interesting to note that
the shear viscosity in a rarefied gas is proportional to the mean free path and hence depends on the coupling constant
as $\lambda^{-2}$.
At large $\lambda T$, $f_\eta$ increases slowly,
and lattice size dependence is found to be small.

Here we compare the present results with those
in the previous work~\cite{Homor} which examined the shear viscosity
near the critical point of the second order phase transition
in the classical $\phi^4$ theory with negative bare mass squared ($m^2=-0.5$).
With the spontaneous symmetry breaking and the thermally induced mass,
the results in Ref.~\cite{Homor} can be regarded as those in the massive cases
except for the region close to the critical temperature.
The shear viscosity in~\cite{Homor} is proportional to the lattice temperature and insensitive to the coupling constant.
Furthermore, the obtained value of it is quite small compared to that in the present work.
At $\lambda T$=1, for example, the shear viscosity divided by the lattice temperature, $\eta/T$, at $m^2 = -0.5$
is smaller than $\eta/T$ at $m^2=0$ by a factor more than 20,
\begin{eqnarray*}
\eta/T|_{\lambda T=1,m^2=0} > 20 \times \eta/T|_{\lambda T=1,m^2=-0.5}.
\end{eqnarray*}
Thus it is found that the shear viscosity of the massless classical
scalar field is more sensitive to the coupling constant and significantly larger than the shear viscosity in the massive case~\cite{Homor}.
The differences from the previous results are caused by the existence of the slow exponential-decay part
in the time-correlation function, which is sensitive to the coupling constant.
While one may consider that thermal fluctuations induce the thermal mass
and the massless property would be lost at finite temperature,
the thermal mass squared induced by thermal fluctuations
is proportional to the coupling constant
and is expected to be small in the weak coupling region,
where the long-time tail of exponential form
appears in the time-correlation function.

\section{Summary}
\label{Sec:Summary}

In this article,
we have studied the shear viscosity of classical scalar fields
on a lattice by using the classical field equation of motion,
the Green-Kubo formula, and the classical field equilibrium configurations.
Using the Green-Kubo formula, we can obtain
the shear viscosity from the time-correlation function
of energy-momentum tensor in equilibrium.
We have adopted the classical field ensemble method,
where the classical field equilibrium configurations are generated
by evolving the system with the artificial Langevin terms.
We have confirmed the equilibrium is reached
from the frequency dependence of the expectation value
of the canonical momentum squared, $\average{|\pi(\bm{k},t)|^2}$.

We have found that the time-correlation function shows oscillatory
behavior around
a slow monotonous decay with an exponential form.
The Fourier spectrum at low frequency is dominated by
the slow decay,
then the shear viscosity is also well evaluated
by the contribution from the exponential decay.

The Fourier spectrum is found to have two components,
a peak at $\omega\simeq 0$ and a bump at $\omega \simeq 4$.
The latter appears also in the free field theories.
By comparison, the former comes from the slow exponential decay part
and suggests that there exists a long-lived excitation mode,
whose pole is on the imaginary axis, $\omega=-i\Gamma$.
Since the damping rate is different from that of the high frequency peak
by two or three orders, this long-lived mode should have the collective nature,
such as the hydrodynamic modes.

We have determined the scaling function for the shear viscosity
of massless classical field,
$f_\eta(\lambda T)=\lambda^2 T\eta(\lambda, T)$,
which is invariant under the scaling transformation.
The scaling function $f_\eta$ is found to take almost constant values
at small $\lambda T$,
which implies that $\eta$ behaves as $1/\lambda^2$ as a function of
the coupling constant and this behavior happens to be
in agreement with the behavior in the rarefied gas,
and increases slowly at large $\lambda T$.
These characteristics of the value and $\lambda$-dependence of $\eta$ are
quite different from those obtained in the previous work~\cite{Homor}
that investigated the shear viscosity near the critical point
of the second order phase transition in the classical $\phi^4$ theory
with negative bare mass squared.
The differences are due to the existence of
the exponential-decay part in the time-correlation function.

While we have discussed the shear viscosity in the massless case ($m=0$),
the dependence on the mass were discussed in Ref.~\cite{Homor}.
It is interesting to examine the role of the slow decay
also in finite mass and negative mass squared cases.
It will be also valuable to discuss the shear viscosity of
classical Yang-Mills field, which is considered to describe the initial
stage of high-energy heavy-ion collisions and has the scaling property
as in the case of the massless classical field theory.
Another important direction to study is to include
the high momentum contributions as well as
the quantum effects.
Equipartition property of classical fields leads to
the Rayleigh-Jeans divergence and nonrenormalizability
in the large cutoff limit, $a^{-1} \to \infty$,
and classical treatment fails to describe high momentum (quantum) dynamics.
In order to evaluate the shear viscosity of quantum systems quantitatively
and to compare it with the perturbative calculation result~\cite{Jeon},
it is desired to take account of low and high momentum modes
with a proper matching procedure
in a nonperturbative way~\cite{Aarts:2001yn,Hatta:2011ky,Ohnishi:1994wn}.

\section*{Acknowledgments}
This work is supported in part by the Grants-in-Aid for Scientific Research
from JSPS (Nos.
15H03663, 
16K05350, 
and 19H01898), 
and by the Yukawa International Program for Quark-hadron Sciences (YIPQS).

\appendix
\section{Time correlation in free scalar theory with Langevin terms}
\label{App:TCFLang}

\begin{figure}[tbhp]
\begin{center}
\includegraphics[width=80mm,bb=0 0 360 252]{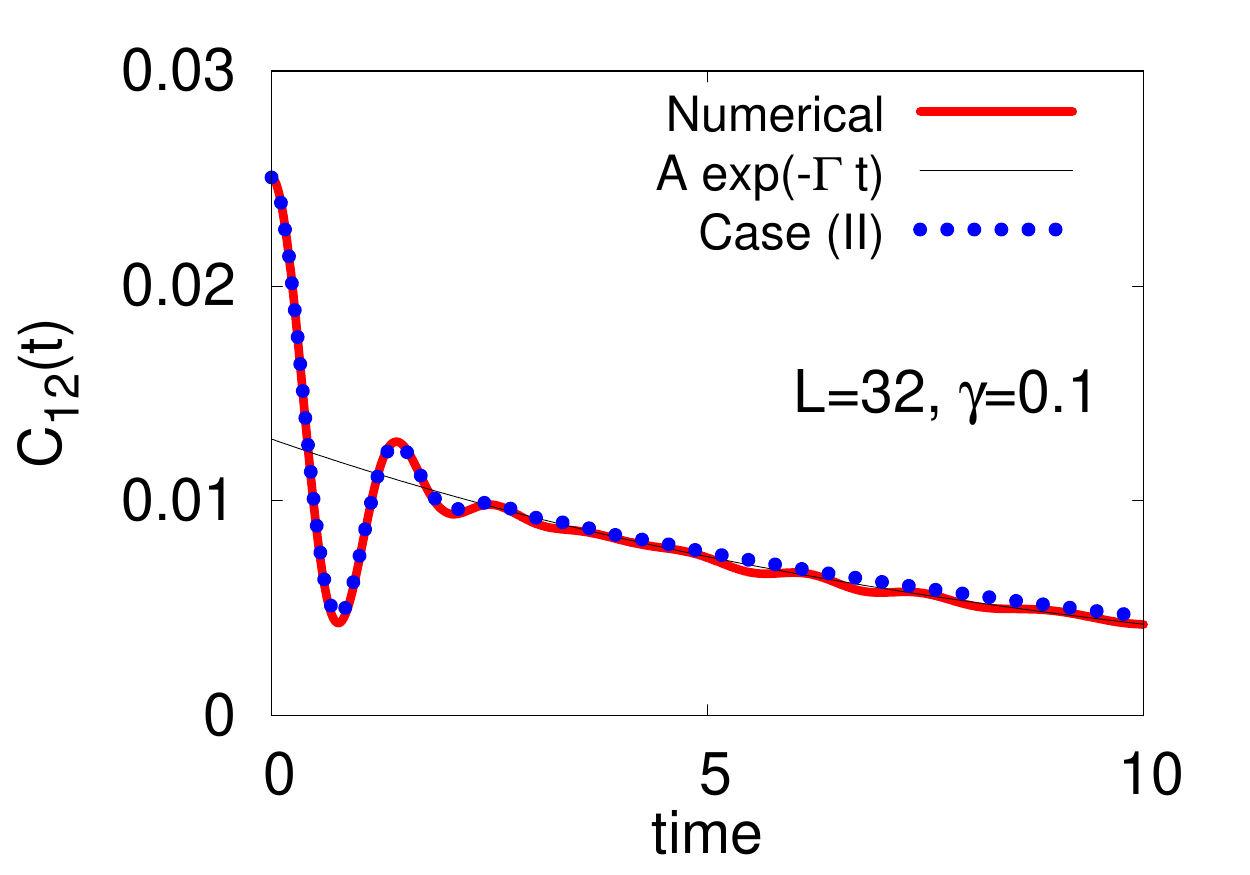}
\end{center}
\caption{The time-correlation function $\TCF(t)$
of the free scalar theory with Langevin terms.
We show the numerical calculation results on a $32^3$ lattice with
$\gamma=0.1$ and $T=1$.
As in the case with interaction,
$\TCF(t)$ shows damped oscillation around the slow exponential decay.
We also show the exponential function fitted to the numerical results (thin solid line) and the expected behavior from analytic calculation
given in Appendix \ref{App:TCFLangED} (dotted line).}
\label{figLT}
\end{figure}

It would be instructive to discuss the thermalization stage
described by the Langevin equation.
We show the time-correlation function of the energy-momentum tensor,
$\TCF(t)=V\average{\tau_{12}(t)\tau_{12}(0)}$,
in the free scalar theory with the Langevin terms in Fig.~\ref{figLT}.
We set the parameters of the Langevin terms
as ($\gamma=0.1, T=1.0$).
The red line shows $\TCF(t)$,
and the black line shows the fitting function.
We find that $\TCF(t)$ dumps exponentially
while it slightly oscillates around the exponential decay part.
We fit $\TCF(t)$ in the exponential function, $\TCF(t)=Ae^{-\Gamma t}$,
where fitting parameter $A$ is the amplitude of the correlation function
and $\Gamma$ is the decay rate, which is the reciprocal of the relaxation time,
$\tau=\Gamma^{-1}$.
After the fitting, we obtain the set of fitting parameter,
\begin{eqnarray}
A &= 0.01238\ \ \ \ \ &+/- 0.00003 (0.2\%),\nonumber\\
\Gamma &= 0.1061\ \ \ \ \ &+/- 0.0003 (0.2\%).\nonumber
\end{eqnarray}
The decay rate is consistent with the diffusion coefficient,
\begin{eqnarray}
\tau^{-1}_{\rm rel}=\Gamma\sim\gamma.
\end{eqnarray}
Then the exponential damp of the correlation function
is found to come from the diffusion term
added in the Langevin equation, Eq.~\eqref{Eq:Langevin}.

\section{Analytic treatment of the time-correlation function with Langevin terms}
\label{App:TCFLangED}

Let us consider the time-correlation function
of the free scalar theory with the Langevin terms
in a model treatment.
For preparation, let us recall the simplest Langevin equation,
\begin{align}
\dot{p}=-\gamma p+R ,
\end{align}
where $R$ is a white noise,
$\average{R(t) R(t')}=2\gamma T\delta(t-t')$.
We can absorb the diffusion term by considering the combination,
$\exp(\gamma t)p(t)$, then it is possible to integrate the equation of motion,
\begin{align}
&\frac{d}{dt}\left(e^{\gamma t} p\right)
=e^{\gamma t} \left(\dot{p}+\gamma p\right)
=e^{\gamma t} R(t)
,\\
&p(t)=e^{-\gamma t}\left(
p(0)+\int_0^t dt' e^{\gamma t'} R(t')
\right)
.
\end{align}
Now let us apply the similar idea to the equation of motion
of the free scalar theory together with the Langevin terms,
which reduces to the Langevin equation for each momentum mode,
\begin{align}
\dot{\pi}_\vk =& -\omega_\vk^2 \phi_\vk - \gamma\pi_\vk + R_\vk
,\\
\dot{\phi}_\vk =& \pi_\vk
,
\end{align}
where $R_\vk$ is a complex white noise,
$\average{R_\vk^*(t) R_{\vk'}(t')}=2\gamma T\delta_{\vk\vk'}\delta(t-t')$.
By combining these equations,
we obtain the equation of one variable $X_\vk=\pi_\vk+i\Omega_\vk \phi_\vk$,
as in the simple Langevin equation,
\begin{align}
\dot{X}_\vk=&i\overline{\Omega}_\vk X_\vk + R_\vk
,\label{Eq:Xk}\\
\Omega_\vk
=&\pm \overline{\omega}_\vk-\frac{i}{2}\gamma
,\\
\overline{\Omega}_\vk
=& \Omega_\vk+i\gamma
= \pm \overline{\omega}_\vk+\frac{i}{2}\gamma
,\\
\overline{\omega}_\vk
=&\sqrt{\omega_\vk^2 - \gamma^2/4}
.
\end{align}
We assume that $\gamma$ is small enough, $\gamma < \omega_\vk/2$,
and $\bomegak$ is real.
Since the first equation \eqref{Eq:Xk} reads
\begin{align}
\frac{d}{dt}\left(e^{-i\bOmega_\vk t}X_\vk\right)=
e^{-i\bOmega_\vk t} R_\vk
,
\end{align}
then we can integrate the right hand side and get,
\begin{align}
X_\vk(t)=&e^{i\overline{\Omega}_\vk t}\left[
X_\vk(0)+ \int_0^t dt' e^{-i\overline{\Omega}_\vk t'} R_\vk(t')
\right]
\ .
\end{align}

In order to obtain $\pi_\vk$ and $\phi_\vk$ separately,
we invoke the two solutions, $\bOmega_\vk=\pm\bomega_\vk-i\gamma/2$,
\begin{align}
(\pik\pm i\bomegak\phik+\frac{\gamma}{2}\phik)_t
=e^{\pm i\bomegak t-\gamma t/2}
\times\left[
(\pik\pm i\bomegak\phik+\frac{\gamma}{2}\phik)_0
+\int_0^t dt' e^{\mp i\bomegak t'+\gamma t'/2} R_\vk(t')
\right]
.
\end{align}
By using these two, we find that $\phik(t)$ is given as the sum of
the diffusion and fluctuation parts,
\begin{align}
\phik(t)=&\phik^{D}(t)+\phik^{F}(t)\ ,\\
\phik^{D}(t)
=&e^{-\gamma t/2}\phik(0)(\cos\bomegak t+\gambar\sin\bomegak t)
+e^{-\gamma t/2}\frac{\pik(0)}{\bomegak}\sin\bomegak t ,\\
\phik^{F}(t)
=&\frac{1}{\bomegak}
\int_0^t dt' R_\vk(t') e^{-\gamma(t-t')/2} \sin\bomegak(t-t') ,
\end{align}
where $\gambar=\gamma/2\bomegak$.

The squared averages are calculated as
\begin{align}
\average{\phik^{F*}(t) \phik^{F}(t)}_\mathrm{eq}
&=\frac{2\gamma T}{\bomegak^2}
\int_0^t dt' e^{-\gamma(t-t')} \sin^2\bomegak(t-t')\nonumber\\
&=\frac{T}{\omegak^2}\left[
1-e^{-\gamma t}\left(1+\gambar^2-\gambar^2\cos2\bomegak t+\gambar\sin2\bomegak t\right)
\right] ,\label{Eq:phiFsq}\\
\average{\phik^{D*}(t) \phik^{D}(t)}_\mathrm{eq}
&=\frac{Te^{-\gamma t}}{\omegak^2}\left[
(\cos\bomegak t+\gambar\sin\bomegak t)^2+(1+\gambar^2)\sin^2\bomegak t
\right] .\label{Eq:phiDsq}
\end{align}
We now confirmed that the equilibrium value
$\average{\phik^*\phik}_\mathrm{eq}=T/\omegak^2$
is obtained at $t \to \infty$ from the fluctuation part,
Eq.~\eqref{Eq:phiFsq}.
Then in Eq.~\eqref{Eq:phiDsq} we have utilized this equilibrium value
and that for $\pik$,
$\average{\pik^*(0)\pik(0)}_\mathrm{eq}=T$.

For the later use, let us consider the average of four field product,
$\average{(\phik^*\phik)^2}$.
We concentrate the fluctuation part, and we find that
\begin{align}
\average{\left[\phik^{F*}(t)\phik^{F}(t)\right]^2}
=2\average{\phik^{F*}(t)\phik^{F}(t)}^2
\to 2\left(\frac{T}{\omegak}\right)^2 .
\end{align}

The factor 2 comes from the two ways of contractions,
and the result looks natural.
Nevertheless, readers may care the product of four white noise terms.
In order to be concrete, we express the white noise
in a finite time step case,
$\int_{t_i}^{t_i+\Delta t} dt R_\vk(t)=\sqrt{\gamma T\Delta t}\xi_i$
and $\xi_i=\xi^R_i+i\xi^I_i$,
where $\xi^R_i$ and $\xi^I_i$ are random numbers in the $i$-th time step
and obey the normal Gaussian distribution,
\begin{align}
\average{\xi^R_i\xi^R_j}=\average{\xi^I_i\xi^I_j}=\delta_{ij},\
\average{\xi^R_i\xi^I_j}=0,
\nonumber\\
\average{\xi^*_i\xi_j}=2\delta_{ij},\
\average{\xi^*_i\xi^*_j}=\average{\xi_i\xi_j}=0.
\end{align}
Then some of the quartic average are given as
\begin{align}
&\average{(\xi^R_i)^4}=\average{(\xi^I_i)^4}=3,\\
&\average{(\xi^*_i\xi_i)^2}=
\average{(\xi^R_i)^4}
+2\average{(\xi^R_i)^2}\average{(\xi^I_i)^2}
+\average{(\xi^I_i)^4}
=8,\\
&\average{\xi^*_i\xi_j\xi^*_k\xi_\ell}
=4(\delta_{ij}\delta_{k\ell}+\delta_{i\ell}\delta_{kj}).
\label{Eq:factorize}
\end{align}
It is interesting to find that there is no additional term
appearing in the case of $i=j=k=\ell$.
By using these relations, the quartic average of
$\phi$ is obtained as
\begin{align}
\average{\left[\phik^{F*}(t)\phik^{F}(t)\right]^2}
=&\int dt_1 dt_2 dt_3 dt_4 f(t_1)f(t_2)f(t_3)f(t_4)
\times
\average{R^*_\vk(t_1)R_\vk(t_2)R^*_\vk(t_3)R_\vk(t_4)}
\nonumber\\
=&(\gamma T\Delta t)^2
\sum_{i,j,k,\ell}
f(t_i)f(t_j)f(t_k)f(t_\ell)
\average{\xi^*_i\xi_j\xi^*_k\xi_\ell}\nonumber\\
=& 2 \left[2\gamma T \Delta t \sum_i f^2(t_i)\right]^2
= 2 \left[2\gamma T \int_0^t dt' f^2(t')\right]^2,
\end{align}
where $f(t')=e^{-\gamma(t-t')/2}\sin^2\bomegak(t-t')/\bomegak$.

We shall now evaluate the time-correlation function.
The spatial average of the energy-momentum tensor ($xy$ element) is given as
\begin{align}
\tau_{12}
=&\frac{1}{V}\sum_{\bm{x}} (\partial_1 \phi) (\partial_2\phi)
=\frac{1}{V}\sum_{\bm{k}} k_1 k_2 \phik^{*} \phik
.\label{Eq:tau12}
\end{align}
The time-correlation function can be evaluated as
\begin{align}
\TCF(t)
=&V\average{\tau_{12}(t)\tau_{12}(0)}_\mathrm{eq}
=\frac{1}{V}\sum_{\bm{k},\bm{k}'} k_1 k_2 k'_1 k'_2
\average{\phik^{*}(t) \phik(t)\phi^{*}_{\vk'}(0) \phi_{\vk'}(0)}_\mathrm{eq}
\nonumber\\
=&\frac{1}{V}\sum_{\bm{k},\bm{k}'} k_1 k_2 k'_1 k'_2 \left[
\average{\phik^{*}(t) \phik(t)}_\mathrm{eq}
\average{\phi^{*}_{\vk'}(0) \phi_{\vk'}(0)}_\mathrm{eq}
+\average{\phik^{*}(t) \phi_{\vk'}(0)}_\mathrm{eq}
\average{\phi^{*}_{\vk'}(0) \phik(t)}_\mathrm{eq}
\right. \nonumber\\&\qquad\qquad\qquad \left.
+\average{\phik^{*}(t) \phi^{*}_{\vk'}(0)}_\mathrm{eq}
\average{\phi_{\vk'}(0) \phik(t)}_\mathrm{eq}
\right]
\nonumber\\
=&\frac{1}{V}\sum_{\bm{k},\bm{k}'} k_1 k_2 k'_1 k'_2
C_\phi(\vk,\vk';t)
=\frac{2}{V}\sum_{\bm{k}} k_1^2 k_2^2 C_\phi(\vk,\vk;t)
,\label{Eq:C12Lang}
\end{align}
where the connected average $C_\phi$ is given as
\begin{align}
C_\phi(\vk,\vk';t)
=&
\average{\phik^{*}(t) \phik(t)\phi^{*}_{\vk'}(0) \phi_{\vk'}(0)}_\mathrm{eq}
-
\average{\phik^{*}(t) \phik(t)}_\mathrm{eq}
\average{\phi^{*}_{\vk'}(0) \phi_{\vk'}(0)}_\mathrm{eq}
\nonumber\\
=&
\average{\phik^{*}(t) \phi_{\vk'}(0)}_\mathrm{eq}
\average{\phi^{*}_{\vk'}(0) \phik(t)}_\mathrm{eq}
+\average{\phik^{*}(t) \phi^{*}_{\vk'}(0)}_\mathrm{eq}
\average{\phi_{\vk'}(0) \phik(t)}_\mathrm{eq}
\nonumber\\
=&
(\delta_{\vk\vk'}+\delta_{\vk,-\vk'})
\times
\average{\phik(t) \phik^{*}(0)}_\mathrm{eq}
\average{\phi_{-\vk}(t) \phi^*_{-\vk}(0)}_\mathrm{eq}
\ ,
\end{align}
where we have used the relation $\phik^*(t)=\phi_{-\vk}(t)$.
The equilibrium average of four field product factorizes
as in the case of the Gaussian noise in Eq.~\eqref{Eq:factorize}.
Then, the contribution from the unconnected part,
$\average{\phik^{*}(t) \phik(t)}_\mathrm{eq}
\average{\phi^{*}_{\vk'}(0) \phi_{\vk'}(0)}_\mathrm{eq}$,
in the third line of Eq.~\eqref{Eq:C12Lang} disappears,
since the equilibrium distribution is isotropic
and the equilibrium average of $\tau_{12}$ in Eq.~\eqref{Eq:tau12} vanishes.
Then, by using the relation $\phi_{-\vk}(t)=\phik^{*}(t)$,
the contributions from $\vk'=-\vk$ as well as those from $\vk'=\vk$
are found to survive,
and we obtain the last line of Eq.~\eqref{Eq:C12Lang}.

We find that only the diffusion part of $\phik(t)$ contribute to $C_\phi$,
since the fluctuation part is independent of $\phik(0)$.
\begin{align}
C_\phi(\vk,\vk;t)
=&e^{-\gamma t}(\cos\bomegak t+\gambar\sin\bomegak t)^2
\times \left[\average{(\phik^{*}\phik)^2}_\mathrm{eq}
-\average{\phik^{*}\phik}_\mathrm{eq}^2\right]
\nonumber\\
=&e^{-\gamma t}
\left(\frac{T}{\omegak^2}\right)^2
(\cos\bomegak t+\gambar\sin\bomegak t)^2
\ .
\end{align}
We have used the equilibrium average of the four field product,
$\average{(\phik^{*}\phik)^2}_\mathrm{eq}=2(T/\omegak^2)^2$.
This value is different from that for the real variable
obeying the Gaussian distribution.
The time-correlation function is now found to be
\begin{align}
\TCF(t)
=&\frac{2e^{-\gamma t}}{V}
\sum_{\bm{k}} k_1^2 k_2^2 \left(\frac{T}{\omegak^2}\right)^2
(\cos\bomegak t+\gambar\sin\bomegak t)^2
.\label{Eq:LangTCF}
\end{align}

\begin{figure}[tbhp]
\begin{center}

\includegraphics[width=80mm,bb=0 0 360 252]{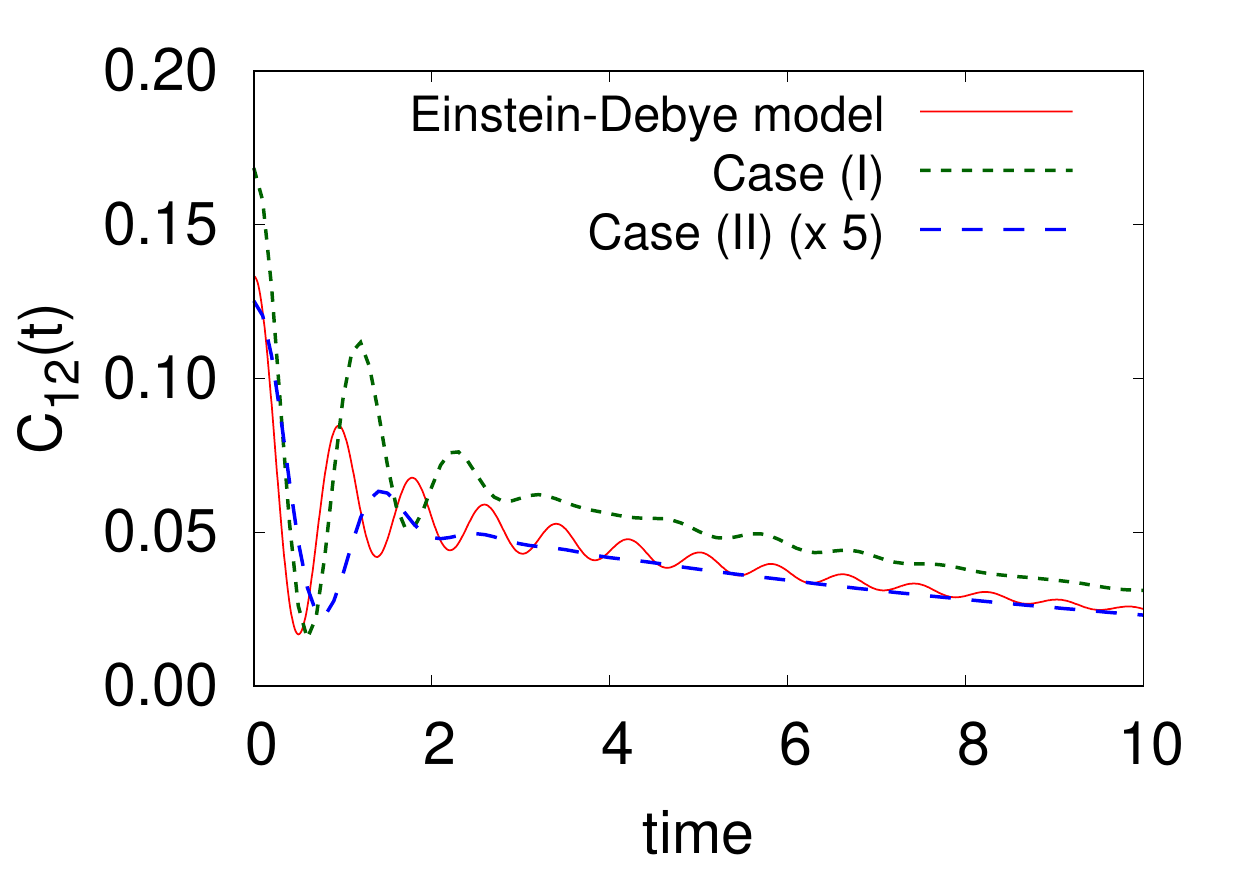}
\end{center}
\caption{The time-correlation function $\TCF(t)$
in free scalar theory with Langevin terms
at $T=1$ and $\gamma=0.1$
In the Einstein-Debye model (solid curve),
we take the cutoff $\Lambda=(6\pi^2)^{1/3}$.
We also show the results with lattice momentum on a $32^3$ lattice.
In the case (I) (green dotted curve) and (II) (blue dashed curve),
we adopt two ways of defining $k_1$ and $k_2$ (see the text).
}
\label{Fig:LangED}
\end{figure}

Let us evaluate $\TCF(t)$ a little more in the Einstein-Debye treatment
of the lattice momentum,
$1/V\times\sum_\vk \to \int^\Lambda d\vk/(2\pi)^3$.
We here ignore the diffusion effect on the oscillating frequency,
$\bomegak \to \omegak \simeq k$.
Then the time-correlation function is evaluated as
\begin{align}
\TCF(t)
\simeq &2T^2e^{-\gamma t}\int^\Lambda\frac{d\vk}{(2\pi)^3}
\frac{k_1^2k_2^2}{k^4}
(\cos kt+\frac{\gamma}{2k}\sin kt)^2
\nonumber\\
=&\frac{T^2\Lambda^3e^{-\gamc x}}{30\pi^2}
\left[f_0(x)+2\gamc f_1(x)+\gamc^2 f_2(x)\right]
,\\
f_0(x)=&\frac13+\frac{1}{x^3}((x^2-2)\sin x+2x\cos x) ,\\
f_1(x)=&\frac{1}{x^2}(\sin x-x\cos x) ,\\
f_2(x)=&1-\frac{\sin x}{x} ,
\end{align}
where $x=2\Lambda t$ and $\gamc=\gamma/2\Lambda$.

In Fig.~\ref{Fig:LangED},
we show $\TCF(t)$ in free scalar theory with Langevin terms.
The qualitative behavior in the numerical results is well explained.
We adopt the cutoff $\Lambda=(6\pi^2)^{1/3}$,
which gives $\int^\Lambda d\vk/(2\pi)^3=1$.
We also show the results using the lattice momentum
in Eq.~\eqref{Eq:LangTCF}.
We consider two cases of the lattice momentum.
Case (I): We define the lattice momentum from the forward difference,
$k_i=2\sin(\pi n_i/L) (n_i=0, 1, \ldots L-1)$.
Case (II): We use $k_i=\sin(2\pi n_i/L)$
expected from the central difference.
The latter results show smaller values of $\TCF(t)$,
and roughly agrees with the numerical calculation also in strength.

The agreement of the numerical and analytical results of $\TCF(t)$
justifies the present treatment, taking the ensemble average
of classically evolved field configurations,
to obtain thermal expectation values
including the shear viscosity in the classical scalar field theory.
In the free scalar theory with Langevin terms,
$\TCF(t)$ is found to show the exponential decay
in the late-time region as in the case with interaction.
This behavior suggests that the interaction term in the $\phi^4$ theory
may be approximately taken into account by the Langevin terms
in the long time region.
It is also suggested that the damped oscillatory behavior
of $\TCF(t)$ in the early stage and the broad bump in the spectral function
can be understood as the contribution of the two-momentum modes,
the contribution from the pile-up of the modes having $\vk$ and $-\vk$.
The frequency of the two-momentum mode depends on $\vk$,
then various two-momentum modes causes the decoherence and damping
of the time-correlation function in the early stage
and the broad bump in the spectral function.

\section{Spectral density}
\label{App:SD}

We here briefly mention the relation of the Fourier spectrum
and the spectral density.
The spectral density $\rho_\mathrm{den}(\omega)$
is defined as the spectral function (Fourier spectrum) of the normal Green's function,
$\average{\left[\tau_{12}(t), \tau_{12}(0)\right]}_{\rm eq}$.
In general, the Fourier spectrum of the product and the spectral density are
related for $\omega\neq0$,
\begin{align}
\rho_\mathrm{den}(\omega)\equiv&\int dt e^{i\omega t}\average{\left[\tau_{12}(t), \tau_{12}(0)\right]}_{\rm eq},\\
I(\omega)\equiv&\int dt e^{i\omega t}\,\frac14\average{\{\tau_{12}(t),\tau_{12}(0)\}}_{\rm eq}
=\frac{1}{2}\coth{\left(\frac{\omega}{2T}\right)}\rho_\mathrm{den}(\omega).
\end{align}
In the low frequency limit,
this relation goes to the simple relation,
\begin{eqnarray*}
I(\omega)&=&\frac{T}{\omega}\rho_\mathrm{den}(\omega) .
\end{eqnarray*}
Thus we can obtain the shear viscosity from the spectral density
in the low frequency limit, $\omega\to0$~\cite{kapusta2006finite},
\begin{eqnarray*}
\eta
=\lim_{\omega\to0}\frac{V I(\omega)}{T}
=\lim_{\omega\to0}\frac{V\rho_\mathrm{den}(\omega)}{\omega}.
\end{eqnarray*}
In the main part of this article,
we have used the Fourier spectrum $\rho(\omega)=VI(\omega)$
instead of the spectral density $\rho_\mathrm{den}(\omega)$
for simplicity.

\end{document}